\documentstyle[12pt,a4]{article}

\newcommand{\wh}{\widehat}
\begin{document}
\begin{titlepage}

\rightline{\tt hep-th/0204031}
\rightline{UT-02-17}

\begin{center}

\vspace{25ex}

{\Large Gauge transformations on a D-brane\\[1.5ex]in Vacuum String Field Theory}

\vspace{10ex}

{\large Yosuke Imamura}%
\footnote{E-mail: {\tt imamura@hep-th.phys.s.u-tokyo.ac.jp}}

\vspace{5ex}

 {\baselineskip=15pt
 {\it Department of Physics,  Faculty of Science, University of Tokyo \\
  Hongo 7-3-1, Bunkyo-ku, Tokyo 113-0033, Japan} 
 }

\vspace{10ex}

\end{center}
\begin{abstract}
We study gauge transformations of the Hata-Kawano vector state
on a D25-brane in the framework of
vacuum string field theory.
We show that among the infinite number of components of the polarization
vector, all the components
except one spacetime vector degree of freedom are gauge freedom,
and give string field gauge transformations reproducing gauge
transformations of the constant modes of the $U(1)$ gauge field.
These gauge transformations can be used to fix the normalization of
the vector field.
We also discuss a difficulty in obtaining
a gauge invariant action of the vector field.
Our arguments rely on the factorization ansatz of
gauge transformations.
\end{abstract}
\end{titlepage}

\newpage

\section{Introduction}
Two kinds of cubic open string field theories,
Witten's cubic string field theory\cite{WittenSFT}
and vacuum string field theory\cite{RSZ0012,RSZ0102,RSZ0105,RSZ0106},
have been recently investigated.
These are considered as two descriptions of one theory covering
both perturbative and nonperturbative vacua.
Although it is expected that these two theories are related to each other
by some (singular) string field redefinition,
derivation of one theory from the other
has not been succeeded.
Now we are at a stage of collecting evidences of the
equivalence of these theories.

In Witten's string field theory,
which describes open strings on D25-branes,
the nonperturbative vacuum state
was constructed by the level truncation method,
and the difference of the energy densities
between perturbative and nonperturbative
vacua was computed numerically with high precision agreement
with the expected D25-brane tension\cite{SZ,leveltr}.
This result was confirmed analytically in the framework of boundary
string field theory\cite{GS}.
Numerical studies of the BRS cohomology at the nonperturbative vacuum
suggested the decoupling of the physical modes\cite{massgen,ET}.

On the other hand, in vacuum string field theory,
D-branes are represented as non-trivial solutions
of the equation of motion.
The action of vacuum string field theory is
\begin{equation}
S[\Phi]=-\frac{1}{g_{\rm v}}\left(\frac{1}{2}\Phi\cdot Q\Phi+\frac{1}{3}\Phi\cdot(\Phi\ast\Phi)\right),
\label{SFTaction}
\end{equation}
where $Q$ is a pure ghost operator and $g_{\rm v}$ is a coupling constant.
The explicit forms of solutions representing D$p$-branes $\Xi_p$
with arbitrary dimensions $p$ are known\cite{KP,RSZ0102}.
By substituting these solutions to the action (\ref{SFTaction})
we can obtain their energy densities ${\cal E}_p$.
Numerical\cite{RSZ0102} and analytical\cite{okuyama}
computations of the ratio ${\cal E}_p/{\cal E}_{p+1}$
showed that it coincides with the expected result $T_p/T_{p+1}=2\pi$.

An attempt to determine the overall factor of the D25-brane energy density
was first done by Hata and Kawano\cite{HK}.
They first constructed a string field solution
representing the tachyon field
on a D-brane as a fluctuation mode of the sliver state $\Xi$,
and computed the three tachyon vertex to fix the ratio between $g_{\rm v}$
in the action (\ref{SFTaction})
and the perturbative coupling constant $g_{\rm str}$.
Disappointingly, the ratio
${\cal E}_{25}/T_{25}=(2\pi)^{25}g_{\rm str}{\cal E}_{25}$
did not coincide with the expected result $1$.
Although the numerical computation was
improved in \cite{hatamori},
the result is close to $2$ rather than $1$.
This unwelcome result was later confirmed analitically\cite{RSZnote}.%
\footnote{After this work was finished
a paper \cite{okawa} appeared where some progress in this popic is made.}

Hata and Kawano also constructed the massless vector field on a D25-brane.
In this paper, we study gauge transformations of
this Hata-Kawano (HK) vector state.
The HK vector state includes a polarization vector
$d^\mu_m$ with a Lorentz index $\mu$ and a level index $m$.
Due to the level index $m$, this has too many components
to represent a vector field.
We will show that, for each spacetime direction,
only one of infinite number of components
corresponds to the $U(1)$ vector field on a D-brane
and all the others are gauge degrees of freedom.
Furthermore, we present a string field gauge transformation reproducing
a gauge transformation of the constant modes of the gauge field
(Wilson lines).
This gauge transformation can be used to fix
the normalization of the gauge field.

Our original motivation was to extract
the coupling constant $g_{\rm str}$
from the kinetic term of the normalized vector field
and to check the coincidence between the energy density of the
sliver state and the expected D25-brane tension.
However, due to a certain problem,
we could not obtain any information of the coupling constant $g_{\rm str}$.
We will discuss this problem in the last section.

We should notice that our arguments in this paper completely
rely on the factorization ansatz of gauge transformations
as will be mentioned in the next section.
At this point, however, we have not shown the existence of
such gauge transformations.
If gauge transformations could not be factorized,
our arguments given below might be spoiled.

Before ending this section,
we define several matrices and vectors
used in this paper.
In the oscillator formalism,
the $3$-string vertex is given by\cite{GJ1,GJ2}
\begin{eqnarray}
&&|V_3\rangle_{\rm mat}=e^E|p_1\rangle|p_2\rangle|p_3\rangle,\\
&&E=\sum_{r,s=0}^3\sum_{m,n=1}^\infty\left(
     -\frac{1}{2}\eta_{\mu\nu}a^{\dagger r\mu}_mV^{rs}_{mn}a^{\dagger s\nu}_n
     -\eta_{\mu\nu}p^{r\mu}V^{rs}_{0n}a^{\dagger s\nu}_n
     -\frac{1}{2}\eta_{\mu\nu}V_{00}p^{r\mu}p^{r\nu}\right).
\end{eqnarray}
The coefficient matrices $V^{rs}$ is called Neumann coefficients.
We define the following matrices and vectors.
\begin{equation}
M_{mn}=(CV^{rr})_{mn},\quad
(M_+)_{mn}=(CV^{rr+1})_{mn},\quad
(M_-)_{mn}=(CV^{rr-1})_{mn},
\end{equation}
\begin{equation}
v_n=V^{rr}_{0n},\quad
(v_+)_n=V^{rr+1}_{0n},\quad
(v_-)_n=V^{rr-1}_{0n},\quad
(n\geq1).
\end{equation}
\begin{equation}
v_1=v_+-v_-,\quad
v_{\pm0}=v_\pm-v.
\end{equation}
where $C_{mn}=\delta_{mn}(-1)^n$ is the twist matrix.
The Lorentz metric $\eta_{\mu\nu}$ and level indices $m,n,\ldots$ are
often suppressed.

%%%%%%%%%%%%%%%%%%%%%%%%%%%%%%%%%%%%%%%%%%%%%%%%%%%%%%%%%
\section{Factorization ansatz of gauge transformations}
In vacuum string field theory,
D-branes are represented as classical solutions of an equation of motion.
The equation of motion obtained from the action (\ref{SFTaction}) is
\begin{equation}
Q\Phi+\Phi\ast\Phi=0.
\label{eom}
\end{equation}
Because operator $Q$ is pure ghost
there exist solutions factored into matter part $\Phi^{\rm mat}$
and ghost part $\Phi^{\rm gh}$:
\begin{equation}
\Phi=\Phi^{\rm mat}\otimes\Phi^{\rm gh}.
\label{factorization}
\end{equation}
In this paper, we always assume this factorization.
$\Phi^{\rm mat}$ and $\Phi^{\rm gh}$ satisfy the following equations.
\begin{eqnarray}
&&Q\Phi^{\rm gh}+\Phi^{\rm gh}\ast\Phi^{\rm gh}=0,\label{ghosteom}\\
&&\Phi^{\rm mat}=\Phi^{\rm mat}\ast\Phi^{\rm mat}.\label{mattereom}
\end{eqnarray}
Although there are several known solutions for the ghost part $\Phi^{\rm gh}$\cite{kishimoto},
we will not specify $\Phi^{\rm gh}$ and concentrate on the matter part.

The action (\ref{SFTaction}) is invariant under the following
gauge transformation.
\begin{equation}
\delta\Phi=Q\Lambda+\Phi\ast\Lambda-\Lambda\ast\Phi.
\end{equation}
For our purpose, `finite' gauge transformations are also convenient.
\begin{equation}
\Phi=U^{-1}\ast QU+U^{-1}\ast\Phi'\ast U.
\label{UDU}
\end{equation}
We ignore the topological term
$\delta S\sim(U^{-1}\ast QU)^3$
because we consider only gauge transformations generated by infinitesimal ones.
The proof of the gauge invariance under (\ref{UDU}) is analogous to that of
three dimensional Chern-Simons theories.
However,
in string field theory,
we need additional assumption,
which is trivial in the Chern-Simons theories.
We have to assume the existence of a string field
${\cal I}$ satisfying
\begin{equation}
Q{\cal I}=0,\quad
{\cal I}\ast\Phi
=\Phi\ast{\cal I}
=\Phi,
\label{Icond}
\end{equation}
for an arbitrary $\Phi$,
and $U^{-1}$ have to be defined from $U$ by $U\ast U^{-1}={\cal I}$.
We are now assuming the factorization (\ref{factorization}).
If we require the gauge transformation
not to break this factorization,
we need to assume that $\cal I$
also factorizes into matter part ${\cal I}^{\rm mat}$
and ghost part ${\cal I}^{\rm gh}$\cite{RSZ0105}.
We can identify ${\cal I}^{\rm mat}$ as the matter identity $I^{\rm mat}$.
The ghost part ${\cal I}^{\rm gh}$ has to satisfy
\begin{equation}
Q{\cal I}^{\rm gh}=0,\quad
{\cal I}^{\rm gh}\ast\Phi^{\rm gh}
=\Phi^{\rm gh}\ast{\cal I}^{\rm gh}
=\Phi^{\rm gh},
\label{Ighcond}
\end{equation}
where $\Phi^{\rm gh}$ is the ghost part of $\Phi$ in (\ref{Icond}).
Naively, the ghost identity $I^{\rm gh}$
satisfies the second equation in (\ref{Ighcond}).
It is known that $QI^{\rm gh}=0$ if we use $\zeta$ function regularization.
Even with other regularization, $QI^{\rm gh}$ is null state, which has
vanishing product with any string field\cite{kishimoto,ko}.
However, precisely speaking, there is no identity field
satisfying $I^{\rm gh}\ast\Phi^{\rm gh}=\Phi^{\rm gh}\ast I^{\rm gh}=\Phi^{\rm gh}$
for any string field $\Phi^{\rm gh}$\cite{kishimoto,ko}.
Even the ghost identity
defined in \cite{GJ2} does not satisfy this condition.
Because of the universality of the ghost factor
this identity condition may unnecessarily be tight.
For our argument below ${\cal I}^{\rm gh}$ does not
have to satisfy (\ref{Ighcond}) for arbitrary $\Phi^{\rm gh}$.
It is sufficient for ${\cal I}^{\rm gh}$ to satisfy
(\ref{Ighcond}) only for the particular $\Phi^{\rm gh}$ adopted as a
universal ghost factor satisfying (\ref{ghosteom}).
In this paper, we simply assume the existence of the state ${\cal I}^{\rm gh}$
satisfying these conditions.

Let us discuss gauge transformations
in the form $U=U^{\rm mat}\otimes{\cal I}^{\rm gh}$.
In this case, the transformation (\ref{UDU}) is
reduced into
\begin{equation}
\Phi'^{\rm mat}=(U^{\rm mat})^{-1}\ast\Phi^{\rm mat}\ast U^{\rm mat}.
\label{fintr}
\end{equation}
We will later show that in vacuum string field theory
the gauge transformations of Wilson lines are reproduced by
the transformation (\ref{fintr}) of the sliver state.
This is different from Witten's string field theory,
in which $U(1)$ gauge transformations are induced by the first term
in (\ref{UDU}) with $Q$ replaced by the BRS charge $Q_B$.

%%%%%%%%%%%%%%%%%%%%%%%%%%%%%%%%%%%%%%%%%%%%%%%%%%
\section{Global gauge transformations of Wilson lines}
From now on, we concentrate only on the matter part of string fields
and we drop the symbol `mat'.

In vacuum string field theory,
D-brane is represented by the sliver state $\Xi$.
It is represented as a squeezed state in oscillator formalizm.\cite{KP,RZ}
\begin{equation}
\Xi\equiv{\cal N}_\Xi e^{-\frac{1}{2}a^{\dagger\mu}CTa^{\dagger\mu}}|0\rangle.
\end{equation}
The matrix $T$ is given by
\begin{equation}
T=\frac{1}{2M}\left(M+1-\sqrt{(1-M)(1+3M)}\right).
\end{equation}
The normalization constant ${\cal N}_\Xi$ is
\begin{equation}
{\cal N}_\Xi=[\det(1-M)(1+T)]^{D/2},
\end{equation}
where $D$ denotes the spacetime dimension $26$.

Now, we suggest the following state as one representing a D-brane with Wilson
line.
\begin{equation}
|\Xi_f\rangle={\cal N}_\Xi(f)e^{-f^\mu_m a^{\dagger\mu}_m}|\Xi\rangle.
\label{Wstate}
\end{equation}
$f^\mu_m$ is a vector representing the amount of the Wilson line
and ${\cal N}_\Xi(f)$ is a normalization factor depending on $f^\mu_m$.
Due to the reality condition, $f^\mu_m$ satisfies $f^{\mu\ast}=-Cf^\mu$.
The fluctuation part of this state reproduces
the zero-momentum limit of the HK vector state\cite{HK}.
We can easily show that this is indeed a projector
if we take the following normalization.
\begin{equation}
{\cal N}_\Xi(f)=
{\cal N}_\Xi
\exp\left(\frac{1}{2}f^\mu\frac{1}{1-T}Cf^\mu\right).
\end{equation}

The HK vector state involves a polarization vector with
infinite number of components.
Similarly, our Wilson line state is parameterized by the vector $f_m^\mu$.
Because of the level index $m$
this vector has infinite number of components.
Of cause only one spacetime vector component of $f_m^\mu$
should be identified with the Wilson line.
Others must be unphysical gauge degrees of freedom.
In this section, we show that we can gauge away all the degrees of
freedom of $f_m^\mu$ except one spacetime vector component.
Let us consider the following gauge transformation specified by
a vector $g_m$.
\begin{equation}
U=I_g={\cal N}_I(g)e^{-\frac{1}{2}a^{\dagger\mu}Ca^{\dagger\mu}-g^\mu a^{\dagger\mu}}|0\rangle,
\label{globalU}
\end{equation}
where ${\cal N}_I(g)$ is a normalization factor depending on the vector $g_m^\mu$.
$g_m^\mu$ is subject to the reality condition $g^{\mu\ast}=-Cg^\mu$.
By demanding $I_0=I_{-g}\ast I_g$ and ${\cal N}(g)={\cal N}(-g)$,
the normalization factor is uniquely determined as
\begin{equation}
{\cal N}_I(g)=\det(1-M)^{D/2}\exp\left(-\frac{1}{4}g^\mu Cg^\mu\right).
\end{equation}
$I_g$ normalized in this way satisfies
\begin{equation}
I_{g_1}\ast I_{g_2}=e^{g_1\times g_2}I_{g_1+g_2},
\label{III}
\end{equation}
where $g_1\times g_2$ is an antisymmetric product of two vectors
$g_1$ and $g_2$ defined by
\begin{equation}
g_1^\mu\times g_2^\mu=g_1^\mu C\frac{M_+-M_-}{2(1-M)}g_2^\mu.
\end{equation}
$I_0$ is the matter identity.
If we put $g_1^\mu=t_1g^\mu$ and $g_2^\mu=t_2g^\mu$ with
some vector $g^\mu$ and real numbers $t_1$ and $t_2$,
(\ref{III}) reduces to $I_{t_1g}\ast I_{t_2g}=I_{(t_1+t_2)g}$.
This implies that $I_g$ is represented as $I_g=\exp\Lambda$
where $\Lambda$ is defined by
\begin{equation}
\Lambda=\left.\frac{dI_{tg}}{dt}\right|_{t=0}=-g^\mu a^{\dagger\mu}|I\rangle.
\label{infgl}
\end{equation}
Products of $\Xi_f$ and $I_g$ are
\begin{equation}
\Xi_f\ast I_g=e^{A(f,g)}\Xi_{f+g(1+T)\rho_R},\quad
I_g\ast\Xi_f=e^{A(f,g)}\Xi_{f+g(1+T)\rho_L},
\label{eq25}
\end{equation}
where
\begin{equation}
A(f,g)=-f^\mu\frac{1}{1-T}Cg^\mu-\frac{1}{4}g^\mu\frac{1+T}{1-T}Cg^\mu.
\label{Afg}
\end{equation}
The matrices $\rho_L$ and $\rho_R$ in (\ref{Afg})
are the half string projection matrices
defined by\cite{halfstring}
\begin{equation}
\rho_L=\frac{M_-+TM_+}{(1-M)(1+T)},\quad
\rho_R=\frac{M_++TM_-}{(1-M)(1+T)}.
\end{equation}
These satisfy the following equations.
\begin{equation}
\rho_L+\rho_R=1,\quad
\rho_L^2=\rho_L,\quad
\rho_R^2=\rho_R,\quad
C\rho_LC=\rho_R.
\end{equation}
Using (\ref{eq25}), the gauge transformation of the Wilson line state $\Xi_f$
by the deformed identity $I_g$ is given as
\begin{equation}
I_{-g}\ast\Xi_f\ast I_g=\Xi_{f'},
\end{equation}
where new vector $f'^\mu_m$ after the transformation is
\begin{equation}
f'=f+g(1+T)(\rho_R-\rho_L).
\label{globalftr}
\end{equation}
Because this is global transformation without momentum,
this does not change the $U(1)$ gauge field on a D-brane.
By this global gauge transformation,
we can gauge away almost degrees of freedom of the vector $f^\mu_m$.
Only components within the kernel of the matrix $(1+T)(\rho_R-\rho_L)$
are invariant under the transformation.
In fact, the matrix has one dimensional kernel.
This can be seen by moving to a basis diagonalizing the matrices
$M$ and $M_\pm$.
In \cite{sas}, diagonalized form of the Neumann matrices
and an associated complete system of eigenvectors
$v_m(\kappa)$ labeled by a real number $\kappa$ are given.
We can expand the vector $f_m^\mu$ by this complete basis as
\begin{equation}
f_m^\mu=\int f^\mu(\kappa)v_m(\kappa)d\kappa.
\end{equation}
Using explicit form of the diagonalized Neumann matrices\cite{sas},
we can rewrite the transformation law (\ref{globalftr}) as
\begin{equation}
f'^\mu(\kappa)=f^\mu(\kappa)+\frac{2\sinh(\pi\kappa/2)}{1+e^{\pi|\kappa|/2}}g^\mu(\kappa).
\label{globalkappa}
\end{equation}
Therefore, the kernel is generated by $\kappa=0$ eigenvector $v_m(0)$
and the Wilson line is identified to $f^\mu(\kappa=0)$
up to the normalization.
This vector is twist odd.
This is consistent with the fact that the vector field on a D-brane is
twist odd.

%%%%%%%%%%%%%%%%%%%%%%%%%%%%%%%%%%%%%%%%%%%%%%%%%%%%%%%%%%%%
\section{Local gauge transformations of Wilson lines}
We have shown that only one spacetime vector degree of freedom of $f^\mu_m$
survives from the global gauge transformation (\ref{globalU}).
In this section, we show that this mode is actually
transformed in a way that $U(1)$ gauge field is transformed
under local gauge transformations.
We consider following string field with non-zero momentum $k^\mu$
as a local gauge transformation.
\begin{equation}
I_{g,k}
=e^{ik^\mu\wh x^\mu}I_g.
\end{equation}
The product of two different string fields of this kind is
\begin{equation}
I_{f_1,p_1}\ast I_{f_2,p_2}
=e^{(f_1,p_1)\times(f_2,p_2)}
I_{f_1+f_2,p_1+p_2}.
\end{equation}
where the anti-symmetric product involving momenta is defined by
\begin{equation}
(f_1^\mu,p_1^\mu)\times(f_2^\mu,p_2^\mu)
=f_1^\mu\times f_2^\mu-v_1\frac{1}{1-M}(p_1^\mu f_2^\mu-p_2^\mu f_1^\mu).
\end{equation}
We also define non-zero momentum states $\Xi_{f,p}$ by
\begin{equation}
\Xi_{f,p}=e^{ip^\mu\wh x^\mu}\Xi_f.
\end{equation}

To compute local gauge transformations generated by $I_{g,k}$,
we use the following formulae:
\begin{equation}
\Xi_{f,p}\ast I_{g,k}=e^{A(f,p,g,k)}\Xi_{f',p+k},\quad
I_{Cg,k}\ast\Xi_{Cf,p}=e^{A(f,p,g,k)}\Xi_{Cf',p+k},
\label{XiIandIXi}
\end{equation}
where $A(f,p,g,k)$ and $f'^\mu$ is given by
\begin{equation}
A(f,p,g,k)=A(f,g)-(p^\mu+k^\mu)v_1\frac{1}{1-M}g^\mu,\quad
f'^\mu=f^\mu-k^\mu v_1\frac{1-T}{1-M}+g^\mu(1+T)\rho_R.
\end{equation}
Combining two equations in (\ref{XiIandIXi}), we easily obtain
the transformation law of Wilson line state as
$I_{-g,-k}\ast\Xi_f\ast I_{g,k}=\Xi_{f'}$
with the transformed vector $f'^\mu_m$ given by
\begin{equation}
f'^\mu=f^\mu+g^\mu(1+T)(\rho_R-\rho_L)-2k^\mu v_1\frac{1-T}{1-M}.
\label{localftr}
\end{equation}
The second term on the right hand side in (\ref{localftr})
is a contribution of global gauge transformation
specified by the parameter $g^\mu_m$ and
this is the same with (\ref{globalftr}).
What is important here is the third term
proportional to the momentum $k^\mu$.
This term is rewritten in the following form
in the diagonalized basis:
\begin{equation}
\delta f^\mu(\kappa)
=-4\sqrt2 k^\mu\frac{\sinh(\pi\kappa/2)}{\kappa(1+\exp(\pi|\kappa|/2))}.
\label{localkappa}
\end{equation}
In contrast to the global transformation (\ref{globalkappa}),
this transformation changes the zero-mode $f^\mu(\kappa=0)$.
Thus, we can identify this transformation to the ordinary
gauge transformation of the $U(1)$ gauge field on a D-brane.
For example, we can define a vector $A_\mu$ by
\begin{equation}
A_\mu=\frac{1}{\sqrt2\pi}f_\mu(\kappa=0).
\label{fA}
\end{equation}
$A_\mu$ represents the constant modes of the $U(1)$ gauge field on a D-brane.
Then, the gauge transformation $U=I_{g,k}$ reproduces the
following gauge transformation of $A_\mu$.
\begin{equation}
A'_\mu=A_\mu-k_\mu.\label{Atr}
\end{equation}
Note that this gauge transformation fixes the normalization
of the gauge field $A_\mu$.
The corresponding normalized zero-momentum HK vector state is given by
\begin{equation}
\Psi=d^\mu_m\left.\frac{\delta\Xi_f}{\delta f^\mu_m}\right|_{f=0}.
\end{equation}
with the polarization vector $d_m^\mu$
satisfying $d^\mu(\kappa=0)=\sqrt2\pi A^\mu$.
(We use $\Psi$ to represent the fluctuation part of the
string field $\Phi$.)

As a more evidence for our Wilson line state indeed to represent Wilson lines,
we can check the shift of the Kaluza-Klein momenta due to Wilson lines.
To obtain a non-trivial result, we should extend the gauge group
to higher rank one.
One way to have higher rank gauge group is to introduce a Chan-Paton factor
by hand.
Let us introduce $U(N)$ Chan-Paton factor indices for
each string field.
We suppose the D-brane solution is diagonal
\begin{equation}
\Xi_{ij}=\Xi_{f_i}\delta_{ij},
\end{equation}
where the vector $(f_i)_m^\mu$ is related to
the Wilson line $A^i_\mu$ for $i$-th $U(1)$ of the Cartan subgroup
via (\ref{fA}).
Let us consider a local gauge transformation in the Cartan subgroup.
The string field representing such a gauge transformation is
\begin{equation}
U_{ij}=\delta_{ij}I_{0,p_i}.
\end{equation}
This transforms the Wilson lines on D-branes as
\begin{equation}
A^i_\mu\rightarrow A^i_\mu-p_\mu^i.
\end{equation}
By this transformation, non-diagonal components of
the fluctuation part of a string field $\Psi_{ij}$ is transformed as
\begin{equation}
\Psi_{ij}\rightarrow
I_{0,-p_i}\ast\Psi_{ij}\ast I_{0,p_j}.
\end{equation}
Due to the momentum conservation,
momentum of $\Psi_{ij}$ is
shifted by $p_j-p_i$.
This represents the momentum shift due to the Wilson lines.

%%%%%%%%%%%%%%%%%%%%%%%%%%%%%%%%%%%%%%%%%%%%%%%%%%%%%%%%
\section{Non-zero momentum modes}
Up to now we have considered only constant modes of the vector field
on a D-brane.
In this section, we discuss the gauge field with non-zero momentum.
The HK vector state
with a momentum $p^\mu$ and a polarization vector $d_\mu^m$ is\cite{HK}
\begin{equation}
\Psi=d^\mu_m a^{\dagger\mu}_m\Xi_{pt,p}.
\label{HKvector}
\end{equation}
where $t_m$ is the following vector defined in \cite{HK}.
\begin{equation}
t=3v(1+3M)^{-1}(1+T).
\end{equation}
The vector state (\ref{HKvector}) is a solution of the linearized
equation of motion:
\begin{equation}
\Psi-\Xi\ast\Psi-\Psi\ast\Xi=0.
\label{lineom}
\end{equation}
In the following we discuss gauge transformation of this mode.
We show that we can again gauge away extra degrees of freedom of
polarization vector $d^\mu_m$
and specify components corresponding to
the physical $U(1)$ gauge field.

Let us consider infinitesimal gauge transformations with a parameter $\Lambda$
given by
\begin{equation}
\Lambda=h^\mu_m\left.\frac{\delta I_{g,k}}{\delta g^\mu_m}\right|_{g=g_0}.
\label{eq50}
\end{equation}
This is non-zero momentum version of the transformation (\ref{infgl}).
The variation of a fluctuation part $\Psi=\Phi-\Xi$ is given by
\begin{equation}
\delta\Psi=\Xi\ast\Lambda-\Lambda\ast\Xi
+{\cal O}(\Psi).
\label{dPhi}
\end{equation}
We consider only the leading term proportional to the background sliver state $\Xi$
and neglect the small contribution from $\Psi$ on the right hand side in (\ref{dPhi}).
To compute this variation, we define
\begin{equation}
X_{g,k}=\Xi\ast I_{g,k}-I_{g,k}\ast\Xi
=e^{A_2}\Xi_{f_2,k}-e^{A_1}\Xi_{f_1,k},
\label{Xdef}
\end{equation}
where
\begin{equation}
f_1^\mu=kv_1\frac{1-T}{1-M}+g^\mu(1+T)\rho_L,\quad
f_2^\mu=-kv_1\frac{1-T}{1-M}+g^\mu(1+T)\rho_R,\quad
\label{f1f2}
\end{equation}
\begin{equation}
A_1=-\frac{1}{4}g^\mu\frac{1+T}{1-T}Cg^\mu+k^\mu v_1\frac{1}{1-M}g^\mu,\quad
A_2=-\frac{1}{4}g^\mu\frac{1+T}{1-T}Cg^\mu-k^\mu v_1\frac{1}{1-M}g^\mu.
\label{A1A2}
\end{equation}
To obtain the last expression in (\ref{Xdef}) we used formulae (\ref{XiIandIXi}).
The variation $\delta\Psi$ is obtained from this $X_{g,k}$ by
\begin{equation}
\delta\Psi=h^\mu_m\left.\frac{\delta X_{g,k}}{\delta g^\mu_m}\right|_{g=g_0}.
\label{dPsifromX}
\end{equation}
General gauge transformations in the form of (\ref{eq50})
do not keep the wave function in the form of (\ref{HKvector}).
Therefore, we consider special gauge transformations which keep the form of (\ref{HKvector}).
In order for the two terms in (\ref{dPhi}) to be added to one term
and to give the variation in the form of (\ref{HKvector}),
both $f_1^\mu$ and $f_2^\mu$ should coincide with $k^\mu t$
when $g^\mu$ is set to $g_0^\mu$.
In fact, by only requiring $f_1^\mu=f_2^\mu$ we can show that
they automatically coincide with $k^\mu t$.
The condition $f_1^\mu=f_2^\mu$ requires the vector $g_0^\mu$ to satisfy
\begin{equation}
g_0^\mu(1+T)(\rho_R-\rho_L)=2k^\mu v_1\frac{1-T}{1-M}.
\end{equation}
Formally, this equation can be solved by multiplying
$(\rho_R-\rho_L)/(1+T)$ on the both sides
and we obtain a solution
\begin{equation}
g_0^\mu=2k^\mu v_1\frac{1-T}{1-M}\frac{\rho_R-\rho_L}{1+T}.
\label{naiveg0}
\end{equation}
This, however, is a dangerous manipulation because $1+T$ has zero eigenvalue.
We will return to this problem later and we now adopt the (singular) solution (\ref{naiveg0}) temporally.
Substituting this solution into (\ref{f1f2}),
we see that $f_1^\mu$ and $f_2^\mu$ actually coincide with $k^\mu t$.
\begin{equation}
f_1^\mu=f_2^\mu=\frac{1}{2}g_0^\mu(1+T)=k^\mu v_1\frac{1-T}{1-M}(\rho_R-\rho_L)=k^\mu t.
\end{equation}
By substituting (\ref{Xdef}) and (\ref{naiveg0}) into (\ref{dPsifromX}),
we obtain
\begin{equation}
\delta\Psi
=[k^\mu t(\rho_R-\rho_L)h^\mu+h^\mu(1+T)(\rho_R-\rho_L)a^{\dagger\mu}]\Xi_{kt,k}
\end{equation}
Because only the twist odd part of
the polarization vector $d^\mu_m$ can include the
$U(1)$ vector modes, let us assume the parameter $h^\mu_m$ of the gauge transformation to be twist even.
Then a transformation low of the polarization vector $d^\mu_m$,
which is identical to that for constant modes, is obtained.
\begin{equation}
\delta d^\mu=h^\mu(1+T)(\rho_R-\rho_L).
\label{eq62}
\end{equation}
Therefore, just as the constant modes,
we claim that the physical vector modes are the $\kappa=0$ components of the polarization vector $d^\mu_m$.

Now, let us return to the problem about the dangerous manipulation used in obtaining $g_0^\mu$ in (\ref{naiveg0}).
One way to regularize this singularity is to use regularized $g_0^\mu$.
We denote the regularized version of $g_0^\mu$ as $g^\mu_{0\rm reg}$.
We take $g^\mu_{0\rm reg}$ such that its $\kappa$ expansion $g_{0\rm reg}^\mu(\kappa)$ does not diverge at $\kappa=0$.
This regularization influents the gauge transformation of $\Psi$.
Especially, the two vectors $f_1^\mu$ and $f_2^\mu$ do not coincide with each other.
The gauge transformation obtained from $g^\mu_{0\rm reg}$ is
\begin{equation}
\delta\Psi=h^\mu(1+T)\left[
\rho_Ra^{\dagger\mu} e^{A_2}\Xi_{f_2,k}
-\rho_La^{\dagger\mu} e^{A_1}\Xi_{f_1,k}
\right],
\label{gtrreg}
\end{equation}
where $f_1^\mu$, $f_2^\mu$, $A_1$ and $A_2$ are those in (\ref{f1f2}) and (\ref{A1A2}) with
$g^\mu$ replaced by $g^\mu_{0\rm reg}$.
(To make the computation simpler, we assumed
the orthogonality $h^\mu k^\mu=0$ and the parallelness $g^\mu_{0\rm reg}\parallel k^\mu$.)
This variation cannot be regarded as a gauge transformation of polarization vector $d^\mu_m$ in
(\ref{HKvector}).
Instead, it can be
regarded as a variation of a
polarization vector of a `regularized' wave function $\Psi_{\rm reg}$.
\begin{equation}
\Psi_{\rm reg}=
d^\mu\rho_Ra^{\dagger\mu}\Xi_{f_2,k}+d^\mu\rho_La^{\dagger\mu}\Xi_{f_1,k}.
\label{Psireg}
\end{equation}
Two vectors $f_1^\mu$ and $f_2^\mu$ in (\ref{Psireg}) are those in (\ref{f1f2})
defined with $g^\mu_{0\rm reg}$, and
this wave function coincides with the original HK vector state (\ref{HKvector})
after taking the limit $g^\mu_{0\rm reg}\rightarrow g^\mu_0$.
For this regularized HK vector state,
the gauge transformation (\ref{gtrreg}) induces the transformation (\ref{eq62})
for the polarization vector $d^\mu_m$ of the reguralized state.
(To obtain this simple expression, we assumed that
$g^\mu_{0\rm reg}$ and $h^\mu$ are twist even and $d^\mu$ is twist odd.
Then $A_1$ is equal to $A_2$ and we can drop the factors $e^{A_1}$ and $e^{A_2}$
by rescaling the gauge transformation parameter $h^\mu$.)
Therefore, we can again claim that only $\kappa=0$ modes of
the polarization vector $d^\mu_m$ is physical degrees of freedom.

%%%%%%%%%%%%%%%%%%%%%%%%%%%%%%%%%%%%%%%%%%%%%%%%%%%%%%%%
\section{Discussions}
Based on the factorization ansatz of gauge transformations,
we have shown that only the $\kappa=0$ components of the polarization vector $d^\mu_m$
are physical degrees of freedom and
all the other components are gauge degrees of freedom.
This implies that if we substitute the wave function (\ref{HKvector}) into the
string field action we must obtain an action depending only on the $\kappa=0$ components
of the polarization vector $d^\mu_m$.
When the gauge field is normalized such that the gauge transformation
does not depend on the coupling constant as (\ref{Atr}),
the coupling constant appears as an overall factor of the action of
the gauge field.
Therefore, if we obtain the kinetic term of the vector field
we can determine the gauge coupling constant as a coefficient of the
kinetic term.
Unfortunately, however, we cannot obtain the gauge invariant action
nor the gauge coupling constant because of a difficulty
we will mention below.

Instead of computing the action of the vector field,
let us consider the linearized equation of motion
 because
the quadratic part of the action is represented
as a BPZ inner product of $\Psi$
and the left hand side of the equation of motion (\ref{lineom}).
We can easily show that the equation of motion
is gauge invariant:
\begin{eqnarray}
&&\delta(\Psi-\Xi\ast\Psi-\Psi\ast\Xi)\nonumber\\
&=&\Xi\ast\Lambda-\Lambda\ast\Xi
-\Xi\ast(\Xi\ast\Lambda-\Lambda\ast\Xi)
-(\Xi\ast\Lambda-\Lambda\ast\Xi)\ast\Xi\nonumber\\
&=&\Xi\ast\Lambda-\Lambda\ast\Xi
-(\Xi\ast\Xi)\ast\Lambda+(\Xi\ast\Lambda)\ast\Xi
-(\Xi\ast\Lambda)\ast\Xi+\Lambda\ast(\Xi\ast\Xi)\nonumber\\
&=&0.
\label{eq63}
\end{eqnarray}
We used associativity of $\ast$-product between
the second expression and the third one in (\ref{eq63}).
On the other hand, if we substitute the solution (\ref{HKvector}) into the equation of motion,
we obtain\cite{HK,hatamori,RSZnote}
\begin{equation}
\Psi-\Xi\ast\Psi-\Psi\ast\Xi=(1-e^{Gp^2/2})\Psi,\quad
G=\log2.
\end{equation}
This is evidently not gauge invariant and depend on all the components of $d^\mu_m$.
This implies that the above `proof'
of the gauge invariance in (\ref{eq63}) is not correct.

In the last section we mentioned that we should regularize the wave function
when we consider the gauge transformation of
non-zero momentum modes.
Namely, more reliable way to compute the gauge invariant equation of motion
and the action is
to use the regularized wave function (\ref{Psireg})
and to take the limit $g^\mu_{0\rm reg}\rightarrow g^\mu_0$ after all the computation.
However, this regularization does not change the situation.
The problem is more serious.

To obtain the gauge transformation $\delta\Psi$,
we need to compute $\ast$-product of the sliver $\Xi$
and the deformed identity state $I_{g,k}$.
$\delta\Psi$ is obtained as a derivative of this product.
Furthermore, to obtain the variation of the equation of motion,
we need to compute $\Xi\ast\Xi\ast I_{g,k}$ and similar products
in different order.
We need the associativity in these products to show the gauge invariance
of the equation of motion.
However, we can easily show that the associativity is broken in
this kind of product.
\begin{equation}
(\Xi\ast\Xi)\ast I_{g,k}\neq\Xi\ast(\Xi\ast I_{g,k}).
\end{equation}
This is the reason why we failed to obtain a gauge invariant equation of motion.
To obtain the coupling constant correctly as a coefficient of the
gauge invariant action,
we should resolve this problem.
At this point, however, we do not have any reliable way to
handle associativity anomaly.

Finally, let discuss other questions which may be interesting.
In the last section, we introduce Chan-Paton factor by hand to
discuss the momentum shift due to the Wilson lines.
More interesting way to construct multiple D-brane is to
use higher rank projector.
In \cite{halfstring}, construction of higher rank projector
as a sum of different rank one projectors is discussed.
They use ``$\ast$-rotation'' to make projectors
orthogonal to each other.
Similarly,
our Wilson line solution may be used to make higher rank projector.
The BPZ inner product and $\ast$-product of two different Wilson line states is
\begin{equation}
\langle\Xi_{f_1}|\Xi_{f_2}\rangle=e^{D(f_1-f_2)}\langle\Xi|\Xi\rangle,\quad
\Xi_{f_1}\ast\Xi_{f_2}=e^{D(f_1-f_2)}\Xi_{f\rho_L+g\rho_R},
\label{WWprod}
\end{equation}
where $D(f)$ is the following function.
\begin{equation}
D(f)=\frac{1}{2}f^\mu\frac{1}{1-T^2}Cf^\mu.
\end{equation}
$1-T^2$ has one zero eigenvalue at $\kappa=0$
and all the other eigenvalues are positive.
We can regularize $D(f)$ by replacing $1-T^2$ by $1-T^2+\epsilon$
with a positive number $\epsilon$.
In the limit $\epsilon\rightarrow+0$,
the exponent $D(f)$ of coefficient of right hand side of (\ref{WWprod})
diverges to minus infinity.
Therefore, if two D-brane have different Wilson lines,
these state are orthogonal to each other.
This implies that $\Xi_f+\Xi_g$ is a projection operator
representing two D-brane solution.

It may be also interesting to study T-duality between Wilson lines on D-branes
and positions of lump solutions\cite{RSZ0102}.
The explicit form of the lump solution is known and we can construct
a solution of a lump at an arbitrary position
by using the shift operator $e^{ia\wh p}$.
One may think that the Wilson line solution studied here would be obtained
from the lump solution by T-duality.
However, the T-duality between them is highly non-trivial.
The gauge fields on D-branes and scalar fluctuation modes of
D-branes have opposite sign of twist.
Thus, T-duality may be represented as an isomorphy of
the $\ast$-algebra exchanging twist even modes $\{a_{2n}\}$ and
twist odd modes $\{a_{2n+1}\}$.

%%%%%%%%%%%%%%%%%%%%%%%%%%%%%%%%%%%%%%%%%%%%%%%%%%%%%%%%%%%%%%%%%%%%%%
\section*{Acknowledgements}
We would like to thank
H.~Hata, I.~Kishimoto, S.~Moriyama and K.~Ohmori
for valuable discussions.
%%%%%%%%%%%%%%%%%%%%%%%%%%%%%%%%%%%%%%%%%%%%%%%%%%%%%%%%%%%%%%%

\end{document}